\documentclass[aps,prl,twocolumn,superscriptaddress,showpacs]{revtex4}
\usepackage{amsfonts,amssymb,amsmath,latexsym,epsfig} 
\begin{document}

[Phys. Rev. Lett. {\bf 96}, 034101 (2006)]

\title{Universality in the synchronization of weighted random networks}

\author{Changsong Zhou}
\affiliation{Institute of Physics, University of Potsdam PF 601553, 14415 Potsdam,  Germany}

\author{Adilson E. Motter}
\affiliation{CNLS and Theoretical Division,  Los Alamos National Laboratory, Los Alamos, NM 87545, USA}
\affiliation{Department of Physics and Astronomy, Northwestern University, Evanston, IL 60208, USA}

\author{J\"{u}rgen Kurths}
\affiliation{Institute of Physics, University of Potsdam PF 601553, 14415 Potsdam,  Germany}

\date{\today}

\begin{abstract}

Realistic networks display not only a complex topological structure, but also 
a heterogeneous distribution of weights in the connection strengths.  Here we
study synchronization in  weighted complex networks and show that the
synchronizability of random networks with large minimum degree is determined by 
two leading parameters: the mean degree and the heterogeneity of the distribution 
of node's intensity, where the intensity of a node, defined as the total strength 
of input connections, is a natural combination of topology and weights. 
Our results provide a possibility for 
the control of synchronization in complex networks 
by the manipulation of few parameters.

\end{abstract}


\maketitle

Over the past few years, the analysis of complex systems from the viewpoint of
networks has become 
an important interdisciplinary issue~\cite{reviews}.
It  has been shown that physical and dynamical processes, such as 
cascading failures~\cite{casc}, epidemic spreadings~\cite{epidemic}, and  network
synchronization~\cite{JJ:2001,Wang:2002,BP:2002,JA:2003,NMLH:2003,MCJ}, are
strongly influenced by the structure of the underlying network.  Previous work
on synchronization has focused mainly on the influence of the topology of the
connections by assuming that the coupling strength is uniform.  
However, synchronization
is influenced not only by the topology, but also by
the strength of the connections~\cite{MCJ}.  Most complex networks where
synchronization is relevant are indeed weighted.  Examples include brain
networks~\cite{brain}, 
networks of coupled populations in the synchronization of
epidemic outbreaks~\cite{city}, and 
technological
networks whose functioning relies on the synchronization of interacting
units~\cite{kor}.  
The distribution of connection weights in real
networks is often highly heterogeneous~\cite{BBPV:2004}.
The study of synchronization in weighted networks is thus of substantial interest.

In this Letter, we address this question in 
random networks with weighted coupling schemes 
motivated by real networks.  Our
main  result is the uncovering of a universal formula that describes
with good approximation the
synchronizability of identical oscillators solely in terms of the mean
degree and the heterogeneity of the node's intensity, irrespective of the
degree distribution
and other topological properties.  The intensity of a node,   
defined as the sum of the strengths of all
input connections of that node,
incorporates both topological and weighted properties and raises as a very
important parameter controlling the synchronizability.
In particular, it follows 
that the synchronizability is significantly enhanced when the heterogeneity of the 
node's intensities is reduced.

The dynamics of a general 
weighted network of $N$ coupled 
identical oscillators is described by   
\begin{eqnarray}
\dot{\bf x}_i&=&{\bf F}({\bf x}_i)+\sigma\mbox{$\sum_{j=1}^{N}$}W_{ij}A_{ij} [{\bf H}({\bf x}_j)-{\bf H}({\bf x}_i)],\label{eq1}\\
         &=&{\bf F}({\bf x}_i)-\sigma\mbox{$\sum_{j=1}^{N}$}G_{ij} {\bf H}({\bf x}_j), \;\;\; i=1,\ldots ,N, 
\label{eq3}
\end{eqnarray}
where ${\bf F}={\bf F}({\bf x})$ governs the dynamics of each individual oscillator, ${\bf H}={\bf H}({\bf x})$ is
the output function, and $\sigma$ is the overall coupling strength.  
Here  $G=(G_{ij})$ is the coupling 
matrix combining  both  topology [adjacency matrix $A=(A_{ij})$] 
and weights [weight matrix  $W=(W_{ij})$, $W_{ij}\ge 0$]:
$G_{ij}=\delta_{ij}S_i-W_{ij}A_{ij}$, 
where $S_i=\sum_{j=1}^N W_{ij}A_{ij}$ denotes the intensity of node $i$.  
The rows of $G$ have  zero sum, and this ensures that the 
completely synchronized state $\{{\bf x}_i={\bf s}, \forall \; i \; | \dot{\bf s}={\bf F}({\bf s})\}$
is  an invariant manifold of Eq.~(\ref{eq3}).  
In this work, we focus on the 
class of weighted networks where $G$ is
diagonalizable and has real eigenvalues.
As it will be shown shortly, in this case the synchronizability 
of the networks can be characterized by  the properties of the eigenvalues, 
without referring to specific  forms of ${\bf F}$ and ${\bf H}$. 
This applies in particular to the important class of networks where $G$
can be written as $G={\cal BC}$ for ${\cal B}$ a nonsingular diagonal matrix and ${\cal C}$ a symmetric,
zero row-sum matrix.
We assume that
matrix $A$ is binary and symmetric, and that the
connection weights and asymmetries are incorporated into $W$.

The linear stability of the synchronized states can be studied by 
diagonalizing the variational equations of Eq.~(\ref{eq3})
into $N$ 
blocks of the form  \cite{msf}
\begin{equation}
\dot{\bf \xi}_l=[D{\bf F}({\bf s})-\sigma \lambda_l D{\bf H}({\bf s}) ]{\bf \xi}_l, \;\;\;
l=1,\ldots , N,
\label{block}
\end{equation}
which are different only  by $\lambda_l$,   
the $l$th eigenvalue of  $G$, ordered as  
$0=\lambda_1\le \lambda_2\cdots\le\lambda_N$.  
The eigenvalues of $G$ are nonnegative because $G_{ii}=-\sum_{j\ne i}G_{ij}\ge 0$
and $\lambda_1=0$ because $\sum_{j}G_{ij}=0$ for all $i$. 
The stability of  ${\bf s}$ is determined by $\lambda_l$ and the 
 master stability function~\cite{msf}, i.e.,  
the largest Lyapunov exponent $\Lambda$ 
of the generic variational equation  
$\dot{\bf \xi}=[D{\bf F}({\bf s})-\epsilon D{\bf H}({\bf s}) ]{\bf \xi}$.  
For many oscillatory dynamical systems~\cite{BP:2002,msf},
$\Lambda$ is negative in a single, finite interval $\epsilon_1<\epsilon<\epsilon_2$,
where the thresholds  $\epsilon_1$ and $\epsilon_2$ are  
determined only by ${\bf F}$, ${\bf H}$, and ${\bf s}$.
The network is thus synchronizable for some 
$\sigma$ iff the condition
$\epsilon_1 <\sigma \lambda_l <\epsilon_2$ is satisfied  so that 
$\Lambda(\sigma \lambda_l)<0$ for all $l\ge 2$.
This is equivalent to the condition
\begin{equation}
R\equiv\lambda_N/\lambda_2<\epsilon_2/\epsilon_1,
\label{R}
\end{equation}
where the eigenratio $R$ depends only on the network structure ($G$), 
and  $\epsilon_2/\epsilon_1$ depends only on the dynamics.
From these, it follows that the smaller the eigenratio $R$ the more
synchronizable the network and vice versa \cite{BP:2002}.
Another measure of synchronizability is
the cost $C$ involved in the couplings of the network~\cite{MCJ}.  When Eq.~(\ref{R})
is satisfied, the synchronized state is linearly stable for
$\sigma> \sigma_{\min}\equiv\epsilon_1/\lambda_2$. The cost $C$ is the total input 
strength of the connections of all nodes at the synchronization threshold: 
$C=\sigma_{\min}\sum_{i,j} W_{ij}A_{ij}=\sigma_{\min}\sum_{i=1}^NS_i$.
The normalized cost
\begin{equation} 
C_0\equiv C/(N\epsilon_1)=\Omega/\lambda_2,
\end{equation}
where $\Omega=\sum_{i=1}^{N}S_i/N$, 
does not depend on the dynamics (${\bf F}$, ${\bf H}$, and ${\bf s}$)
and can be used as a complementary parameter of synchronizability~\cite{MCJ}.
Here we characterize the synchronizability of 
the networks
using both the eigenratio $R$ and the cost $C_0$.

Previous work 
has obtained bounds
for the eigenvalues of 
{\it unweighted}  networks ($W_{ij}=1$)
(see \cite{pb} for a review).
Such bounds, however, are not tight and may provide limited information about 
the actual synchronizability of complex networks.
Here we aim at obtaining a more quantitative approximation of the 
synchronizability for a 
class of weighted random networks with real spectra, which includes  
unweighted networks as a special case. 
Our analysis is based on the combination of a
mean field approximation and new graph spectral results.

First, in random networks with $k_{\min}\gg1$, close to 
a synchronized state
Eq.~(\ref{eq1}) can be approximated as 
\begin{equation}
\dot{\bf x}_i={\bf F}({\bf x}_i)+\sigma (S_i/k_i) 
\mbox{$\sum_{j=1}^{N}$}A_{ij}[{\bf H}({\bf x}_j)-{\bf H}({\bf x}_i)].
\label{app_1}
\end{equation}
The reason is that each oscillator $j$ receives  signals from  a large  and random sample of
other  oscillators in the  network and ${\bf x}_j$ is not affected directly
by the individual  output weights $W_{ij}$. Consequently, we may assume that $W_{ij}$
and ${\bf H}({\bf x}_j)$ are statistically  uncorrelated and that
$\sum_{j=1}^{N}W_{ij} A_{ij} {\bf H}({\bf x}_j) \approx 
(1/k_i)\sum_{j=1}^{N}W_{ij}A_{ij}  \sum_{j=1}^{N}A_{ij} {\bf H}({\bf x}_j)=S_i\bar {\bf H}_i$  if $k_i\gg 1$ 
\cite{eff_k}.                       
Here 
$\bar{\bf H}_i=(1/k_i)\sum_{j=1}^{N}A_{ij} {\bf H}({\bf x}_j)$  is the local mean field.

Now, if the network is  sufficiently random, 
the local mean field
 $\bar {\bf H}_i$  can be approximated by the global
mean field of the network,  $\bar {\bf H}_i\approx \bar{\bf H}=
(1/N)\sum_{j=1}^{N}{\bf H}({\bf x}_j)$.
Moreover, 
close to the synchronized state ${\bf s}$,
we may assume $\bar {\bf H}_i\approx {\bf H} ({\bf s})$,
and the system is approximated as
\begin{equation}
\dot{\bf x}_i={\bf F}({\bf x}_i)
+\sigma S_i[ {\bf H}({\bf s})-{\bf H}({\bf x}_i)], \;\; i=1,\dots,N,
\label{global}
\end{equation}
indicating that the oscillators are decoupled and forced by
a common  oscillator $\dot {\bf s}= {\bf F}({\bf s})$
with forcing strength
proportional to  the intensity $S_i$.
The variational equations of Eq.~(\ref{global}) have  the same form 
of Eq.~(\ref{block}), except that $\lambda_l$ is replaced by $S_l$.   
If there exists some $\sigma$ satisfying  $\epsilon_1< \sigma S_l <\epsilon_2$
for  all $l$,
then all the oscillators
are synchronizable by the common driving ${\bf H}({\bf s})$,
corresponding to a complete synchronization
of the whole network.
These observations  suggest that the eigenratio and the cost can be
approximated as 
\begin{equation}
R\approx S_{\max}/S_{\min}, \;\;\; C_0\approx \Omega/S_{\min},
\label{R_app}
\end{equation} 
where $S_{\min}$,  $S_{\max}$, and $\Omega$ are the minimum, maximum and
mean intensities, respectively. 

Next we present tight bounds for the above approximation. 
Eq.~(\ref{app_1}) means that 
the coupling matrix $G$ is  replaced by the  new matrix $G^a=(G^a_{ij})$, 
with $G^a_{ij}=\frac{S_i}{k_i}(\delta_{ij}k_i-A_{ij})$.   
$G^a$ can be written as $G^a=S\hat{G}=SD^{-1}(D-A)$, where 
$S=(\delta_{ij}S_i)$ and $D=(\delta_{ij}k_i)$ are the diagonal matrices of
intensities  and degrees, respectively, and $\hat{G}$ is the normalized Laplacian 
matrix~\cite{CLV:2003}. Importantly, now the contributions 
from the topology  and weight structure are separated and 
accounted by $\hat{G}$ and $S$, respectively. 
We can show that 
the largest and smallest nonzero eigenvalues of 
matrix $G^a$ are bounded by the eigenvalues $\mu_l$ of $\hat{G}$  as
\begin{equation}
S_{\min}\mu_2 c \le \lambda_2 \le  S_{\min} c', \;\;\;
S_{\max}\le \lambda_N \le S_{\max}\mu_N, \label{bound_2}
\end{equation}
where $c$ and $c'$ can be approximated by $1$ for most large complex 
networks of interest, such as realistic scale-free networks (SFNs).
The proof is involved and long, and the details will be presented 
elsewhere.
If the network 
is sufficiently random, the  spectrum of $\hat{G}$
tends to the semicircle law for large networks with arbitrary expected
degrees~\cite{CLV:2003}, provided that $k_{\min}\gg \sqrt{K}$, and
$\max \{1-\mu_2,\mu_N-1\}= [1+o(1)]\frac{2}{\sqrt{K}}$ for
$k_{\min}\gg \sqrt{K}\ln^3 N$, where $K$ is the mean degree.  
From these, it follows that
\begin{equation}
\mu_2\approx 1-2/\sqrt{K}, \;\; \mu_N \approx 1+2/\sqrt{K},  
\label{mu_2N}
\end{equation}
which we find to provide
a good approximation   under the  weaker condition $k_{\min}\gg 1$,
regardless of the degree distribution.
From Eqs.~(\ref{bound_2}) and (\ref{mu_2N}), we have 
the following  
approximations for the bounds of $R$ and $C_0$:
\begin{eqnarray}
\frac{S_{\max}}{S_{\min}} 
\le&R& \le 
\frac{S_{\max}}{S_{\min}} \frac{1+{2}/{\sqrt{K}}}{1-{2}/{\sqrt{K}}}, 
\label{R_bound}\\ 
\frac{\Omega}{S_{\min}} 
\le&C_0& \le
\frac{\Omega}{S_{\min}} \frac{1}{1-{2}/{\sqrt{K}}}. \label{C_bound}
\end{eqnarray}
For the case of unweighted networks ($S_i=k_i$, $\Omega=K$), the bounds in Eq.~(\ref{R_bound}) 
are much tighter than those reviewed in Ref.~\cite{pb}.

The bounds in Eqs. (\ref{R_bound}) and (\ref{C_bound}) show that the 
contribution of the network topology is mainly accounted by the mean 
degree $K$. Therefore, for a given $K$, the synchronizability of random 
networks with large $k_{\min}$ is expected to be well approximated by 
the following  universal formula:      
\begin{equation}
R=A_R \frac{S_{\max}}{S_{\min}}, \;\;
C_0=A_C \frac{\Omega}{S_{\min}},
\label{fit_all}
\end{equation}
where the pre-factors  $A_R$ and $A_C$ are expected to be close to 1.  
In the case of matrix $G^a$ with uniform intensity ($S_i=1 \; \forall i$), they are 
given by the upper bounds,   $A_R=\frac{1+{2}/{\sqrt{K}}}{1-{2}/{\sqrt{K}}}$ 
and $A_C=\frac{1}{1-{2}/{\sqrt{K}}}$, and
$A_R\to 1$ and $A_C \to 1$  in the limit $K\to \infty$.
Formula (\ref{fit_all}) is consistent with the approximation in Eq.~(\ref{R_app})
and indicates that the synchronizability of these networks is primarily
determined by the heterogeneity of the intensities.

Our numerical  results on various weighted and unweighted networks have 
confirmed this  universal formula.  
First we consider the following 
weighted  coupling scheme:
\begin{equation}
W_{ij}=S_i/k_i,
\label{weight}
\end{equation}
in which the intensities $S_i$ 
follow an arbitrary distribution not necessarily correlated
with the degrees. 
In this case, Eq.~(\ref{app_1}) and Eq.~(\ref{eq1}) are identical 
and $G^a=G$.  
This weighted coupling scheme  includes many
previously studied systems as special cases.
If  $S_i=k_i \;  \forall i$, it corresponds to
the widely studied case of unweighed
networks~\cite{BP:2002,Wang:2002,NMLH:2003}. 
In the case of fully uniform intensity ($S_i=1 \; \forall i$),
it accommodates a number of previous studies 
about synchronization of
coupled maps~\cite{JJ:2001,JA:2003}.
The weighted scheme studied in~\cite{MCJ}, $W_{ij}=k_i^\theta$,
is another special case of Eq.~(\ref{weight}) where $S_i=k_i^{1+\theta}$.

We have applied  the weighted scheme  to 
various network models: 
{\it (i) Growing SFNs with aging}~\cite{DM:2000}.  
Starting with $2m+1$ fully connected nodes,  at each time step we 
connect a new node to $m$ existing nodes according 
to the probability $\Pi_i \sim k_i \tau_i^{-\alpha}$, where $\tau_i$ is the age 
of the node.
The minimum degree is then $k_{\min}=m$ and the mean degree is $K=2m$.
For the aging exponent $-\infty<\alpha \leq 0$, this growing rule generates SFNs with a power-law
tail  $P(k)\sim k^{-\gamma}$ and scaling exponent in the interval $2<\gamma\leq 3$ \cite{DM:2000},
as in most real SFNs.
For $\alpha=0$, we recover the usual BA model~\cite{sf}, which has $\gamma=3$.
{\it (ii) Random SFNs}~\cite{NSW:2001}. 
Each node is assigned
to have a number $k_i \ge k_{\min}$ of ``half-links'' according to the
distribution 
$P(k)\sim k^{-\gamma}$.
The network is generated by
randomly connecting these half-links to form links, prohibiting self- and
repeated links. 
{\it (iii) $K$-regular random networks.}  Each node  is  randomly connected to $K$ other
nodes. 

We present results for  two different distributions of intensity $S_i$
which are uncorrelated with the distribution of degree $k_i$:
(1) a  uniform distribution  in $[S_{\min}, S_{\max}]$; and
(2) a  power-law  distribution, $P(S)~\sim S^{-\Gamma}$, $S\ge S_{\min}$, where
$S_{\min}$ is a positive number. 
Consistently with the prediction of the universal formula, 
if $k_{\min}\gg 1$,  
the  eigenratio $R$   collapses  into a single curve  for a given $K$  
when  plotted as a function of $S_{\max}/S_{\min}$  [Fig.~1(a)], 
irrespective of  the distributions of $k_i$ and $S_i$.    
The same happens for the cost $C_0$ as a function of $\Omega/S_{\min}$ [Fig.~1(b)]. 
The behavior of the fitting parameters $A_R$ and $A_C$ is shown in the insets of Fig.~1.
For uniform intensity, they are very close to the upper bounds. They
approach very quickly 1  when the intensities become more heterogeneous
($S_{\max}/S_{\min}> 3$). Therefore, Eqs.~(\ref{fit_all}) with $A_R=A_C=1$ [Fig.~1, solid lines] 
provide a good approximation of the synchronizability for any large $K$ 
if the  intensities are not very homogeneous.

\begin{figure}[pt]
\begin{center}
\epsfig{figure=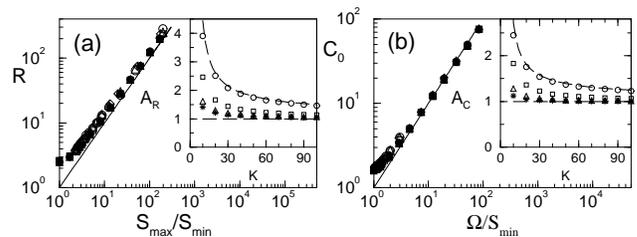,width=8.2cm}
\caption{(a) $R$ as a  function of $S_{\max}/S_{\min}$ and (b)
$C_0$ as a function of $\Omega/S_{\min}$, averaged over 50 realizations of the networks.
Filled symbols: uniform distribution of $S_i \in [S_{\min},S_{\max}]$.
Open symbols:  power-law  distribution of $S_i$, $P(S)\sim S^{-\Gamma}$
for $2.5 \le\Gamma\le 10$.
Different symbols are for networks with  different topologies:
BA growing SFNs  ($\circ$), growing SFNs with aging exponent $\alpha=-3$ ($\Box$), 
 random SFNs with $\gamma=3$ ($\diamond$),
and $K$-regular random networks ($\triangle$). The number of nodes is $N=2^{10}$ and the mean degree is $K=20$.
Insets of (a) and (b): $A_R$ and $A_C$ as functions of $K$ for 
$S_{\max}/S_{\min}=1$ $(\circ)$, $2$ $(\Box)$, $10$ $(\triangle)$,  and  $100$ $(\ast)$, obtained 
with  uniform distribution of $S_i$ in $K$-regular networks.   
The dashed lines are  the  bounds.  
Solid lines in (a) and (b): Eqs.~(\ref{fit_all}) with $A_R=A_C=1$. 
}
\end{center}
\end{figure}

In more realistic networks, 
including scientific collaboration
networks~\cite{BBPV:2004}, metabolic networks~\cite{MAB:2004},
and airport networks~\cite{BBPV:2004,MAB:2004}, it has been shown  that 
the weight $W_{ij}$ of a connection between nodes $i$ and $j$ is
strongly correlated with the product of the corresponding degrees
as
$\langle W_{ij}\rangle \sim (k_ik_j)^\theta$.  Here $\theta$ 
depends on the
specific network under study.  
Motivated by these observations, we analyze  
the  weighted coupling \cite{MAB:2004}:
\begin{equation}
W_{ij}=(k_ik_j)^\theta, 
\label{real_weight}
\end{equation} 
where the weights are defined for the connections of a given network topology
and $\theta$ is a tunable parameter.
$\theta$ controls  the heterogeneity of  the intensity $S_i$ and  the  
correlation between $S_i$ and $k_i$,  
since $S_i=k_i^{1+\theta} \langle k_j^\theta\rangle_i$,
where $\langle k_j^\theta\rangle_i =(1/k_i)\sum k_j^\theta$ is  approximately 
constant for $k_i\gg 1$ when the degree correlations can be neglected. 
Variations of  $\theta$ have significant impact on the synchronizability of SFNs [Fig.~2].
However, as shown in the insets of Fig.~2 for various networks 
and $\theta$ values, $R$ and $C_0$  collapse again to the 
universal 
curves when regarded as functions of $S_{\max}/S_{\min}$ and $\Omega/S_{\min}$,  respectively. 
 
We  emphasize that these results
also hold for unweighted random networks.
As shown in Fig.~3 for unweighted SFNs,  
the eigenratio $R$ as a function of the network size $N$
collapses to a single universal curve  
when plotted against $S_{\max}/S_{\min}$ \cite{note}.          

In addition to heterogeneous degrees and weights, many real networks also
display high clustering~\cite{sw} and nontrivial correlation of
degrees~\cite{N:2002}.
The clustering and degree correlations are negligible in
random SFNs and $K$-regular random networks, but we find that
they are significant in growing SFNs for $\alpha<0$.
The results in Figs.~1-3 show that 
the universality holds  
without significant dependences on these topological properties.

\begin{figure}[pt]
\begin{center}
\epsfig{figure=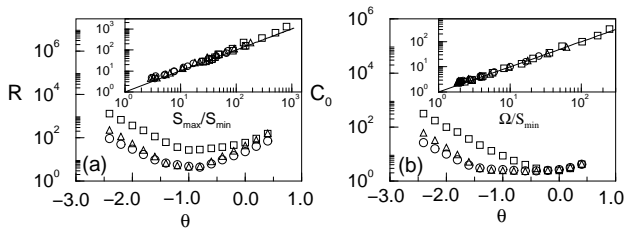,width=8.2cm}
\caption{ (a) Eigenratio $R$ and (b) cost $C_0$ as functions of $\theta$ 
for the BA growing SFNs  ($\circ$), growing SFNs with aging exponent $\alpha=-3$ ($\Box$)
and random SFNs with $\gamma=3$ ($\triangle$). 
Each symbol is an average over 50 realizations of the networks with  $K=20$ and $N=2^{10}$.
Inset of (a): the same data for  $R$ as a function of $S_{\max}/S_{\min}$. 
Inset of (b): the same data for  $C_0$ as a function of $\Omega/S_{\min}$. 
Solid lines:  Eqs.~(\ref{fit_all}) with $A_R=A_C=1$.
}
\end{center}
\end{figure}

Eqs.~(\ref{fit_all})
also provide a meaningful approximation 
for networks which are not fully random. 
For example, consider small-world networks~\cite{sw}  where 
a regular ring of $N$ ($=2^{10}$) nodes, each  connected to $K$ ($=20$)  
nearest neighbors, is rewired with a probability $p$ for each link.
We find that $R$ and $C_0$ collapse to the universal curves  
even when the networks are dominated by local connections, e.g.
for $p= 0.3$,  if the intensities are very heterogeneous ($S_{\max}/S_{\min}\ge 10$).     
For networks with $k_{\min} \sim 1$, the synchronizability is still 
strongly dependent on
$S_{\max}/S_{\min}$ and $K$, although it shows 
additional dependences on the details of the distributions of $S_i$ and $k_i$
and on other  topological properties.

\begin{figure}[pt]
\begin{center}
\epsfig{figure=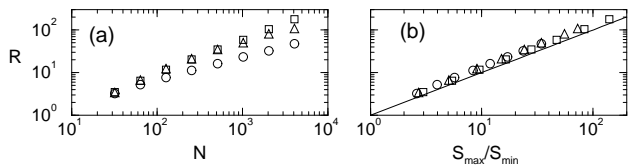,width=8.2cm}
\caption{(a) $R$ as a function of the system size $N$ for unweighted networks ($\theta=0$).   
(b) The same data for  $R$ as a function of $S_{\max}/S_{\min}$.
Solid line: Eq.~(\ref{fit_all}) with $A_R=1$.
The other parameters and legends  are the same as in Fig.~2(a)
}
\end{center}
\end{figure}

In summary, we have shown that the synchronizability of sufficiently random
networks with minimum degree $k_{\min}\gg 1$ is 
universally dominantly determined by the
mean degree $K$ and the heterogeneity of the intensities $S_i$.  This
universality applies to a general class of large networks where the
heterogeneity of $S_i$ is due to either the distribution of degrees, as in
unweighted SFNs, or the distribution of connection weights, as in weighted
$K$-regular networks, or a combination of both, as expected in most realistic
networks,   such as in the airport network~\cite{BBPV:2004}, which underlies
the synchronization of epidemic outbreaks 
\cite{city}. 
In particular, formula (\ref{fit_all})
explains why synchronizability is
improved when the heterogeneity of $S_i$ is reduced, which can be useful for
network design and control of synchronization.   

C.\ S.\ Z.\ and J.\ K.\ were  supported by the VW Foundation and SFB 555.
A.\ E.\ M.\ was supported by DOE under Contact No.\ W-7405-ENG-36.


\begin{references}
\bibitem{eff_k}
The condition for the approximation
to be valid is that $s_i/(\sqrt{k_i}a_i) \ll 1$, where $a_i$ and $s_i$ are the average and standard deviation of the
$k_i\gg 1$ values of $W_{ij}$, respectively. 
Numerically, we find that the approximation often remains valid under milder conditions.
In the case that the total strength of the $k_i$ connections is dominated by $k_i^{*}< k_i$ links,
and the other connections have much weaker strengths, we assume that $k_i$ is replaced by $k_i^{*}$ throughout the paper.
\bibitem{note} In this case, 
$\Omega/S_{\min}=K/k_{\min}$,
so that $C_0$ 
is almost constant against $N$, and is not shown.
\end{references}
\end{document}